\documentclass[twocolumn]{customarticle}  

\usepackage{graphicx}
\usepackage{txfonts}
\usepackage{hyperref} %
\usepackage{textcomp} %
\usepackage{rotating}
\usepackage{threeparttable} %
\usepackage{multicol} %
\usepackage{multirow}
\usepackage{ulem} 
\usepackage{microtype}
\usepackage{natbib}
\hypersetup{
  colorlinks=true,
  linkcolor=blue,
  citecolor=blue,
  urlcolor=blue
}
\usepackage{xcolor} %

\title{Modelling the eruptive young stellar object Re 50 N IRS 1 with ProDiMo}

\author{Andreas Postel}

\begin{document}

\twocolumn[
    \begin{@twocolumnfalse}
        \begin{center}

    \maketitle
    \begin{abstract}
    \end{abstract}

    \end{center}
    \end{@twocolumnfalse}
]
\section{Introduction}
Episodic accretion is recognized to play a central role in the accretion history of young stars. One of the most outstanding examples of episodic accretion are
FUors, named after the archetype FU Orionis, which are low-mass young stars that experience strong optical outbursts that can last several decades. FU Orionis had
its outburst in 1936, reaching an increase of 6 mag in the B-band at the peak, and has since then been slowly fading \citep{2000ApJ...531.1028K}.
A few dozens of objects with similar behaviour have been identified since the outburst of FU Ori (\citealt{1996ARA&A..34..207H}; \citealt{2010vaoa.conf...19R}). There have also been detections of some candidates (\citealt{2007ApJ...658..487Q}; \citealt{2010ATel.2801....1S}; \citealt{2012ApJ...748L...5R}; \citealt{2012ApJ...756...99F};
\citealt{2013AN....334...53F}) that have not been observed in the pre-outburst state, which however show typical characteristics of FUors.
FUors appear to be predominantly protostars which still accrete from their parental envelopes or are in the early T-Tauri star phase with
only a remnant envelope.

Characteristic of FUors are the very strong outbursts, with a luminosity change from 4 to 6 mag which occur on timescales of days to weeks and last for several decades. Another type of eruptive low-mass stars is
characterised by weaker (3 to 5 mag) and shorter outbursts which however can occur more frequently \citep{1989ESOC...33..233H}. EX Lupi is the archetype of this second class, which is accordingly called EXors. There are indications that these objects are more evolved than FUors, but the transition between the two classes may be smooth \citep{Audard2014}. The formation of massive stars also diplays signatures of episodic accretion \citep[e.g.][]{2017sfcc.confE..15D,2017arXiv171002320M}, suggesting that this phenomenon is somewhat universal among star formation.

A number of investigations at the infrared and the sub-millimeter regimes during the recent years finally allowed accessing of the dust emission and lines of FUors and EXors in this wavelength range, which provided insights into the circumstellar disk and possible envelopes around these objects (\citealt{Lorenzetti2005}, \citealt{Green2013}).
It was found to be extremely common that young stars show activity in the form of significant IR variability. The inner regions of FUors and EXors, which are of high interest to reveal the origin of an outburst, are challenging for single-dish observations as these phenomena cannot be proven unambiguously to occur in the inner disk alone. Recent studies with the Atacama Large Millimeter/submillimeter Array (ALMA) and Karl G. Jansky Very Large Array (JVLA) opened a new window for investigations of the outburst objects with very high spatial resolution and allowed the observation of hot inner disks \citep{2017A&A...602A..19L}, density waves in the disks \citep{2016Sci...353.1519P}, outflows of the objects (\citealt{2017MNRAS.468.3266R,2017MNRAS.466.3519R}), and shifts of the snow line during outburst \citep{2016Natur.535..258C}.

The origin of accretion bursts is still unclear and could be caused by viscous-thermal and/or magnetorotational instabilities that form in the inner disk, tidal effects
caused by close companions or close flybys of external stars, or accretion of gaseous clumps in gravitationally unstable disks (see \citealt{Audard2014} for a review).
We aim to investigate the properties of FUors and EXors by analysing observational data \citep{Postel_2019}, along with numerical hydrodynamics simulations of protostellar disks \citep{2015ApJ...805..115V}, stellar evolution models of outbursting stars \citep{2019MNRAS.484..146E} and thermo-chemical models of star--disk systems in the outburst state \citep{Rab2017}. The goal of our project is to get a better understanding of the outburst processes and their respective origin.

In this paper, we focus on the FUor-like object Re 50 N IRS 1 and present the results of a ProDiMo model (ProDiMo is a radiation thermo-chemical code to model protoplanetary disks) during a re-brightening phase. Interestingly, \citealt{2015ApJ...805...54C} proposed that the object is going through a dust clearing phase in the optical and near-IR. Focussing on mid-/far-IR and sub-mm data, we analyse what Herschel and Spitzer observations can tell us in the constraints of the model. ProDiMo has been extended to handle envelope structures \citep{Rab2017}, allowing it to not only model T-Tauri stars but also the earlier evolutionary stages like FUors and EXors.

The main questions we aim to answer in this paper are how the gas lines of the object can be modelled and where they come from, as well as whether the envelope plays a special role in the scenario of Re 50 N IRS 1 and if we need episodic accretion to explain the observational data.

\begin{figure}
    \includegraphics[width=0.5\textwidth, trim={0cm 0cm -0.2cm 0}]{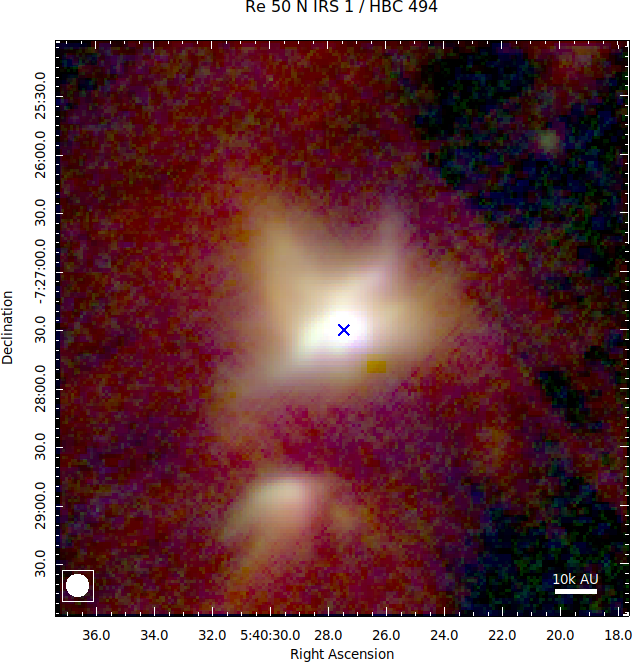}
        \caption{
            \footnotesize
           Image of Re 50 N IRS 1 in PACS. The colors refer to the three different photometry channels of PACS, centered at 70 \textmu m (blue), 100 \textmu m (green) and 160 \textmu m. The target coordinates of the FUor are marked with a blue cross. The beam of the longest used wavelength is illustrated by the circle in the lower left corner. We used a log scaling and removed values below zero. The spectra originate from the inner 39 arcsec.
        }
        \label{image_Re50}
\end{figure}

\subsection{Re 50 N IRS 1}
Re 50 N IRS 1 / HBC 494 is a known FUor-like object \citep{Audard2014} in the Orion A cloud, which is 414$\pm$7 pc \citep{Menten_2007} away from us, with a wide spread \citep{Grossschedl_2018} based on Galactic longitude. The object has a bolometric luminosity of 90 $\mathrm{L}_\sun$, based on the SED of our recent work in \citet{Postel_2019}. It was observed in the past with several instruments, covering the SED from the optical to the mm range with photometry and spectra, including Herschel (Sep 2013), Spitzer (7 Nov 2008) and ALMA (Aug 2015, see \citealt{2017MNRAS.466.3519R}) observations. Our Herschel data shows many emission lines of Re 50 N IRS 1, mainly from CO from low to high excitation temperatures, which makes the object a promising candidate for detailed modelling. The object shows strong silicate absorption in the bending and stretching bands and CO$_2$ absorption. Re 50 N IRS 1 is known for its very wide outflow, detected by ALMA, and strong diffuse emission in two Herschel instruments (PACS/SPIRE: 60-671 \textmu m) extending to in the northeast and to the west, see Fig. \ref{image_Re50}.

\begin{table*}[htb]
    \caption{Main parameters of the best fitting ProDiMo model.}
    \label{parameters}
    \centering
    \begin{tabular}{l|c|c|c}
    \hline\hline
        Quantity        & Symbol        &    Value                &     Reference    \\
        \hline
        stellar mass    & $M_\star$        & 0.5 $\mathrm{M}_\sun$&    this work              \\
        central source eff. temp. & $T_\star$    & 6000 K         &    this work              \\
        central source luminosity    & $L_\star$    & 90 $\mathrm{L}_\sun$ &    this work, based on \cite{Postel_2019}              \\
        amount of UV radiation    & fUV & 3\%                     &    this work              \\
        \hline
        disk gas mass   & $M_{\mathrm{disk}}$    & 0.02 $\mathrm{M}_\sun$           &   \cite{2018MNRAS.474.4347C}               \\
        disk accretion rate & $\dot{M}$ & 6 $\times 10^{-4}$ $\mathrm{M}_\sun$ yr$^{-1}$ &  this work          \\
        disk inner radius & $R_{\mathrm{in}}$& 1.0 au                      &    assumed              \\
        disk tapering-off radius & $R_{\mathrm{tap}}$ & 100 au             &   \cite{2018MNRAS.474.4347C}               \\
        inclination & incl & 15.3°                                &   this work                \\
        column dens. pow. ind. & $\epsilon$ & 1                   &    assumed              \\
        reference scale height & $H(100 au)$ & 10 au              &    assumed              \\
        flaring power index & $\beta$ & 1.1                       &    assumed              \\
        \hline
        envelope mass & $M_{\mathrm{env}}$ & $\approx$ 2.57 $\mathrm{M}_\sun$                 &   this work               \\
        mass infall rate & $\dot{M}_{\mathrm{if}}$ & 1.35 $\times 10^{-5}$ $\mathrm{M}_\sun$ yr$^{-1}$ &  fitted          \\
        outer radius & $R_{\mathrm{out}}$ & 15000 au                       &   calc., based on \cite{Postel_2019}               \\
        cavity opening angle & $\beta_{\mathrm{cav}}$ & 15°                &   \cite{2017MNRAS.466.3519R}               \\
        \hline
        dust to gas mass ratio & $\delta$ & 0.01                  &   \cite{2018MNRAS.474.4347C}               \\
        min. dust particle radius & $a_{\mathrm{min}}$ & 0.05 \textmu m    &    assumed              \\
        max. dust particle radius & $a_{\mathrm{max}}$ & 3000 \textmu m    &    assumed              \\
        dust size dist. power ind. & $a_{\mathrm{pow}}$ & 3.5              &    assumed              \\
        dust composition & Mg$_{0.7}$ Fe$_{0.3}$ SiO$_3$ & 60\%   &    assumed              \\
        (volume fractions) & amorph. carbon & 20\%                &    assumed              \\
                          & vacuum & 20\%                         &    assumed              \\
        \hline
        visual extinction & $A_V$ & 3.1                           &   this work               \\
        distance & d & 414 pc                                     &   \citet{Menten_2007}               \\
    \hline
    \end{tabular}
        \tablefoot{The model with no UV field has fUV=0 and the model with no disk accretion has $\dot{M}$=0. The assumed parameters were fixed during the fitting process.}
\end{table*}

\section{Methods}
The radiation thermo-chemical code, ProDiMo (see \citealt{Kamp2017}, \citealt{Thi2011}, \citealt{Woitke2009b}, \citealt{Woitke2016}), solves the temperature, chemical abundances and radiation field in a self-consistent way under constraints of
axisymmetric distribution of gas and dust in two dimensions. The protostar is presented by a point source with corresponding luminosity in the center of the structure, surrounded by a protoplanetary disk and spherical envelope with outflow cavities.
The code is used to derive atomic and molecular abundances in different regions of the modelled object, and calculate the spectral lines (both in emission and absorption) and the spectral energy distribution (SED). This model was already used in \citealt{Rab2017}, where a detailed explanation, about how the disk and envelope structure is built, is explained. In \citealt{2019ApJ...877...21W} this model was used to fit APEX data with 3 lines but without Herschel data. More details about the disk accretion heating are shown in Appendix \ref{appendix:diskaccretionheating}.

\section{Modelling}
The procedure to find a good matching model can be separated in three steps. The first phase was to check the literature for already determined parameters. After that we matched the SED with ProDiMo. Once this was done, we matched the gas emission lines.
We investigated several configurations in ProDiMo to probe a large parameter space. In Table \ref{parameters}, we provide the parameters of the model
that we consider fits best the Re 50 N IRS 1 gas lines and SED (mainly the Herschel one, see below).

    \subsection{Literature}
    \cite{2018MNRAS.474.4347C} derived the dust mass from one ALMA continuum image. Assuming a gas to dust mass ration of 100, they estimated the disk mass,
    which are both used in this work beside the tapering parameter.
    We use a reference scale height of the disk of
    10 au which is slightly above the margin of \cite{2018MNRAS.474.4347C}. \cite{2017MNRAS.466.3519R} delivered a measurement with ALMA of the opening angle of
    the cavity. Other parameters like the effective temperature of the central source, the mass infall rate and the disk accretion
    rate as well as the amount of the UV radiation were derived in our analysis which is described later in more detail. Based on photometry of Herschel (PACS)
    and Hubble, we estimate the outer radius of the system of Re 50 N IRS 1 at up to 15,000 au. Within this distance, there is no significant overlap of
    the emission of the object and the nearby reflection nebula. We assume that the line flux measured by Herschel from this region is not contaminated by another
    source for the modelling.

    \subsection{SED matching} \label{SED_fitting}
    The second step in our analysis was to match the broad-band SED, while focussing on the Herschel and Spitzer data. With the SED matching, we obtained better
    initial conditions for changing the model parameters and fit the gas emission lines. This provides constrains for the disk and envelope dust structure.
    We varied several parameters while we discuss here the ones with the biggest impact on the SED. Other parameters that have been explored but are not discussed in detail are the luminosity of the central source, effective temperature of the central source and the cavity opening angle, as their impact on the SED was very limited.
    We focussed on Herschel and
    Spitzer photometry and spectra between 10-700 \textmu m. The Herschel data were
    observed in 2013 \citep{Postel_2019}, before the object was found to go through a brightening phase \citep{2015ApJ...805...54C}. The data in the optical and near-IR,
    as well as the major part of the photometry in the sub-mm and mm are older (1988-2001). We take here the approach that such data, while informative, do not necessarily
    correspond to the SED state during the Herschel observation, and therefore we do not optimize the parameters to match them.
    The Herschel data is therefore used as our benchmark.
    
    The first parameter we started with
    is the mass infall rate onto the envelope, as this has the biggest impact on the emission in the far-IR/sub-mm. In the model of the rotating envelope, this parameter has no heating effect on the structure.
    The change
    of the SED is shown in Fig. \ref{SED_grid2}. For the mass infall rate
    for our best model (blue curve, $\dot{M}\approx10^{-5}$ $\mathrm{M}_\sun$ yr$^{-1}$) the far-IR/sub-mm data of Herschel ar
    matched very well, which was
    the main goal of this step, but there is an extreme variation in the wavelength range of the Spitzer spectrum and for shorter wavelengths when $M_{if}$ is varied. The next step was to match the
    mid-IR, for which the inclination of the object had a major impact, which left the far-IR/sub-mm range mostly unchanged. We show in
    Fig. \ref{SED_grid10} the corresponding SED for a variation of the inclination. While the model predicts higher flux between the Herschel PACS
    spectrum and the Spitzer spectrum, a solution for the short wavelength domain was covered within the parameter space at an inclination between 10° and 16.25°, which we refine to 15.3°.
    As the object is embedded and stellar parameters are mostly unknown, we also ran a grid from 0.1-0.8 $\mathrm{M}_\sun$ where we find minor differences in the SED. The
    range is reasonable for such an object, and we eventually use a stellar mass of 0.5 $\mathrm{M}_\sun$, however there is potential for large difference to the actual protostar. We further assume that there is extinction along the line of sight that is not part of the model. This extinction is set by the $A_V$ parameter. We find that we need some additional extinction to match the short-wavelength regime.
    The result of the entire SED matching
    process can be seen in Fig. \ref{SED}, together with observational data and the spectrum of the protostar without absorption caused by the envelope and disk.
        
    \subsection{Gas emission lines}
    The next step of our analysis was to match the gas emission lines detected with Herschel while keeping the SED untouched.
    Fig. \ref{lines_grid18} shows the emission line fluxes from Herschel together with the ProDiMo modelled fluxes based on the SED-matching model, with a variation of the stellar luminosity. The emission
    lines at long wavelengths ($\gtrapprox$ 300 \textmu m) are matched quite well by the model, but the shorter wavelength lines, being orders of magnitudes too low, are not
    properly reproduced. Since the short wavelength CO lines, originating from
    high J transitions, and the oxygen emission are caused by material which is much hotter than the material which emits the long wavelength lines, we increased the effective temperature of
    the central source of the model to heat some material further. The result can be seen in Fig. \ref{lines_grid19}. The impact was negligible. Looking at the
    gas temperature, we found that the CO and oxygen of the disk was simply not getting warm enough, and even raising the effective temperature of the central source
    to 12000 K did not help. 
    
    We therefore investigated another parameter and introduced a UV field around the central source, assuming that accretion shocks
    near the protostar would create a significant UV radiation field, indicated by the high level of [\ion{O}{i}] emission, that would in turn heat the disk and envelope temperature sufficiently high. The temperature
    increase was, however, concentrated to the surface of the disk and did not lead to more emission of high J CO lines at some point but photodissociated the CO. We
    therefore introduced disk accretion to have a disk-internal heating process in addition to the UV field. Both processes, which affect only the gas, are disabled in the default settings
    but appear to be a necessity for this particular object.
        
\section{Results}
In Fig. \ref{emissionlines}, we show three different models. Our best match result of the SED and lines and two other models models (Mdot=0, fUV=0) to compare the effects of disk accretion and UV field of the central source: The other two models have either disabled disk accretion but an UV field or disk accretion but no UV field. Neither the UV field nor the disk accretion heating mechanism are sufficient to raise the temperature of the disk high enough to reproduce the
observational measurements. We believe therefore, that the model with both disk mass accretion and UV field is a better model to match the gas emission lines and keep the fit of the SED intact.

\begin{figure}
    \includegraphics[width=0.5\textwidth, trim={0cm 0cm 0cm 0}]{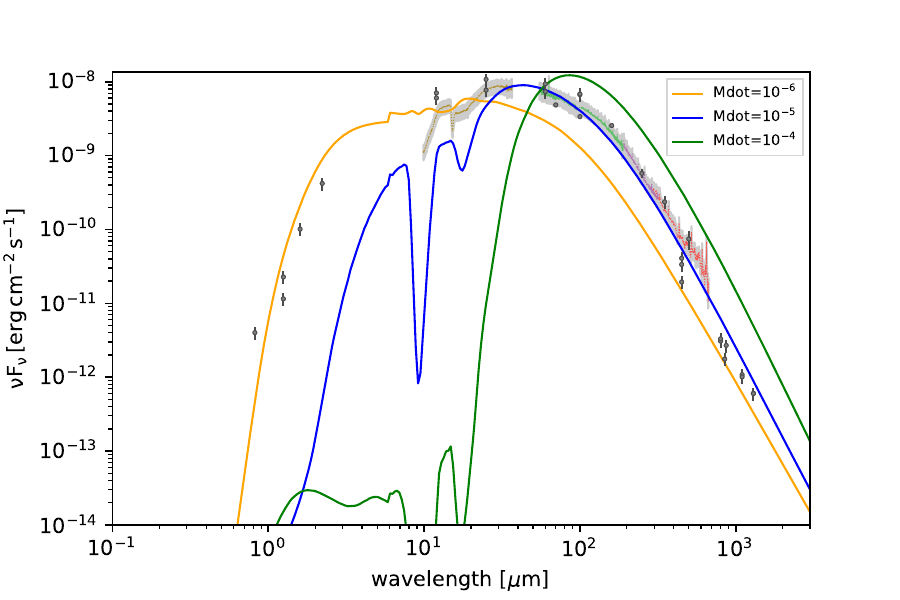}
        \caption{
            \footnotesize
            Change of the SED with different mass infall rates ($10^{-6}-10^{-4}$ $\mathrm{M}_\sun$ yr$^{-1}$) for the envelope. The goal of this grid was to match the
            far-IR/sub-mm data. The Herschel and Spitzer spectra which worked as a benchmark for the SED matching are shown in colors (Spitzer: brown;
            Herschel PACS: green; Herschel SPIRE SSW: purple; Herschel SPIRE SLW: red)
        }
        \label{SED_grid2}
\end{figure}

\begin{figure}
    \includegraphics[width=0.5\textwidth, trim={0cm 0cm 0cm 0}]{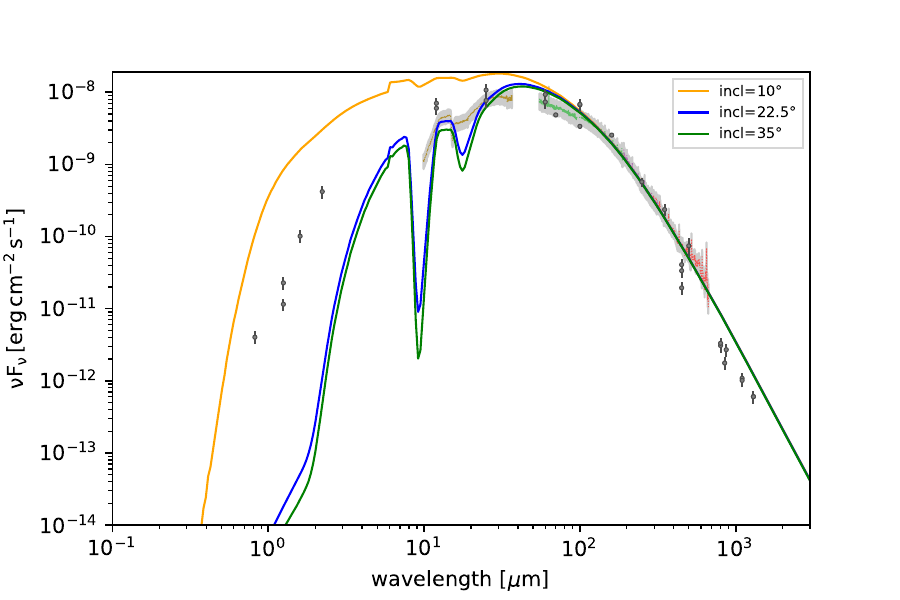}
        \caption{
            \footnotesize
            Change of the SED with inclination angles, varying from 10-35°. The goal of this grid was to match the mid-IR data. The colour code is the same as in
            Fig. \ref{SED_grid2}.
        }
        \label{SED_grid10}
\end{figure}

\begin{figure}
    \includegraphics[width=0.5\textwidth, trim={0cm 0cm 0cm 0}]{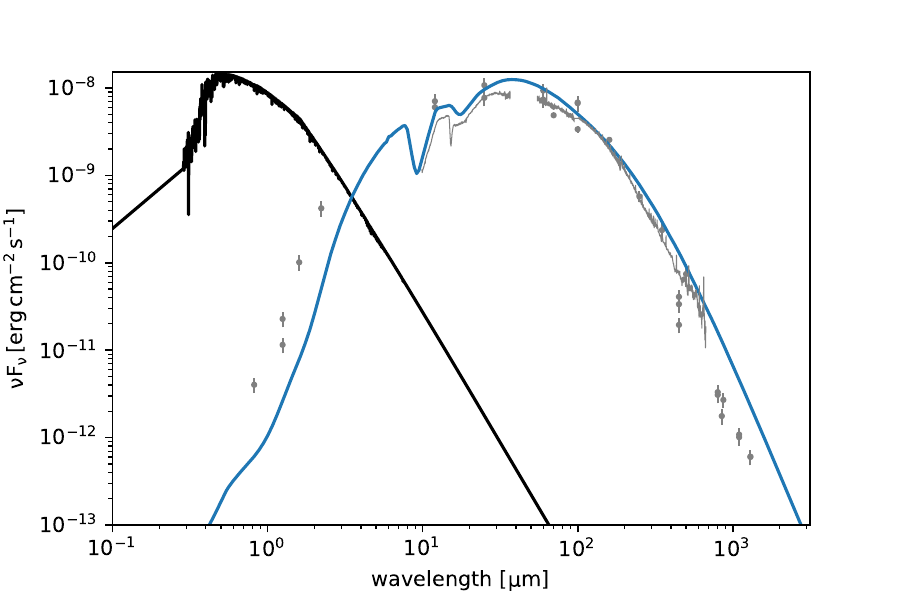}
        \caption{
            \footnotesize
            SED of the model (blue) and the observational data of photometry and spectroscopy (grey). The spectrum of the central source is shown in black.
        }
        \label{SED}
\end{figure}

\begin{figure*}
    \includegraphics[width=1.0\textwidth, trim={0cm 0cm 0cm 0}]{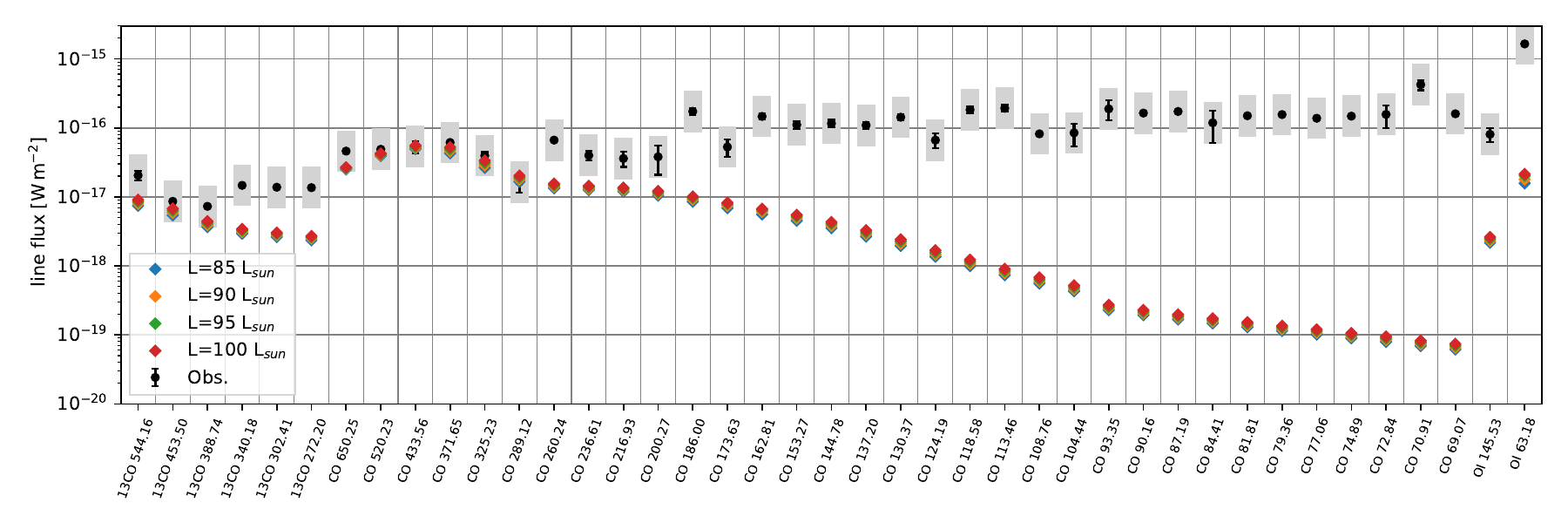}
        \caption{
            \footnotesize
            Emission lines after the SED matching was performed. Here we show the effect of different luminosities of the central source in comparison to
            the Herschel line fluxes. The numbers next to the molecules/atoms refer to the wavelength of the respective lines in micrometer. The grey boxes represent a factor of 3 margin around the observational data.
            The change of the luminosity did not have a strong impact on the lines.
        }
        \label{lines_grid18}
\end{figure*}

\begin{figure*}
    \includegraphics[width=1.0\textwidth, trim={0cm 0cm 0cm 0}]{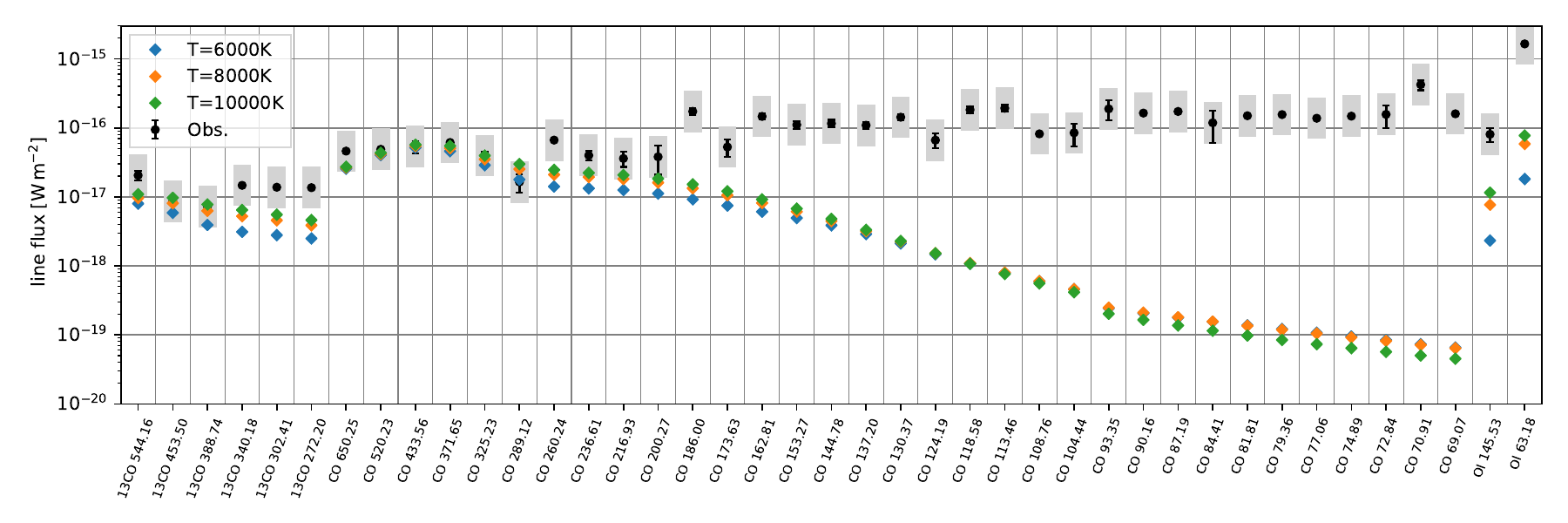}
        \caption{
            \footnotesize
            Same as Fig. \ref{lines_grid18}, but for different central source effective temperatures, ranging from 6000 to 10000 K.  
        }
        \label{lines_grid19}
\end{figure*}

\begin{figure*}
    \includegraphics[width=1.0\textwidth, trim={0cm 0cm 0cm 0}]{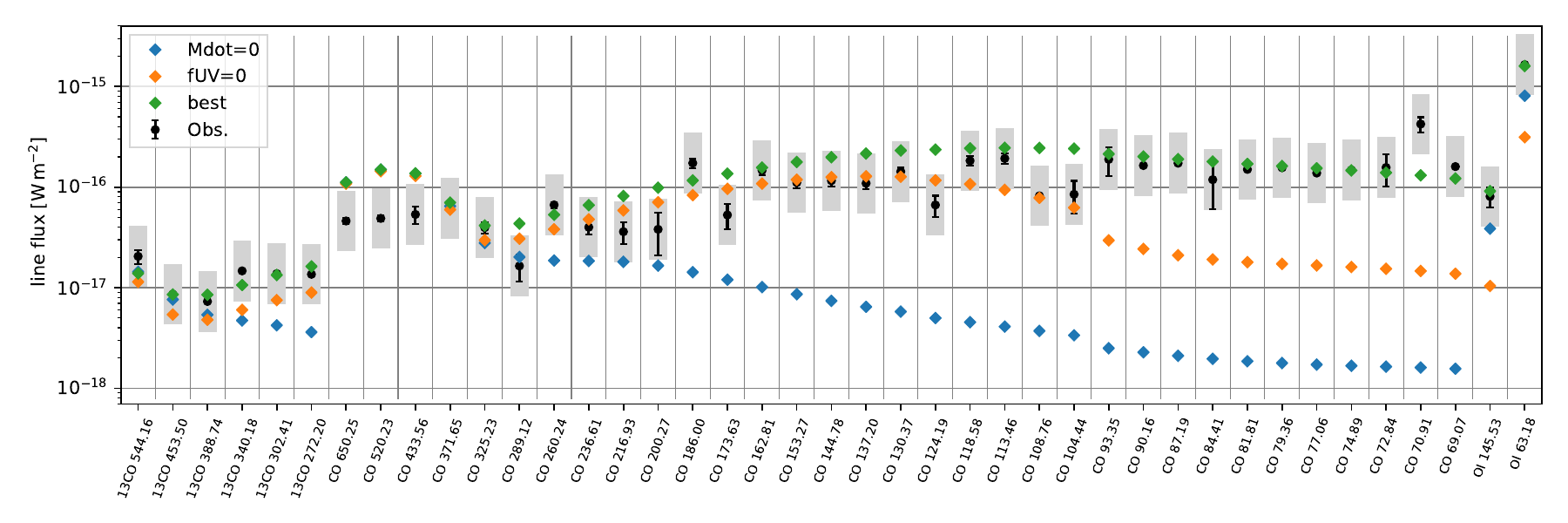}
        \caption{
            \footnotesize
            Line fluxes of the best model (green diamonds) compared to a model with disabled
            disk accretion ($\mathrm{\dot{M}=0}$, fUV = 1\%, blue diamonds) and with disabled UV field ($\mathrm{\dot{M}}=4.2\times 10^{-4}$ $\mathrm{M}_\sun$ yr$^{-1}$, fUV = 0, orange diamonds). The best model uses both, disk accretion ($\mathrm{\dot{M}}=6\times 10^{-4}$ $\mathrm{M}_\sun$ yr$^{-1}$ and the UV field (fUV = 3\%) in combination.
        }
        \label{emissionlines}
\end{figure*}

    \subsection{Spectral properties and surrounding emission} \label{subsection:SEDs_and_photometry}
        The matching of the SED (Fig. \ref{SED}) with ProDiMo, mainly the Herschel and Spitzer data, was precise enough to recreate the results from the observations
        reasonably well. At long wavelengths there is a difference to some
        ground based observations. Also, there is a gap to observational data in the range of 800 nm to 3 \textmu m. These observations were performed before the
        re-brightening of the object. 
        The difference to the
        ground-based observations in the (sub-)mm range are explained by our model by a change of the mass infall rate of the envelope by about one order of magnitude (see Fig. \ref{SED_grid2}).

\begin{figure*}[!ht]
    \includegraphics[width=0.5\textwidth, trim={0cm 0cm 0cm 0}]{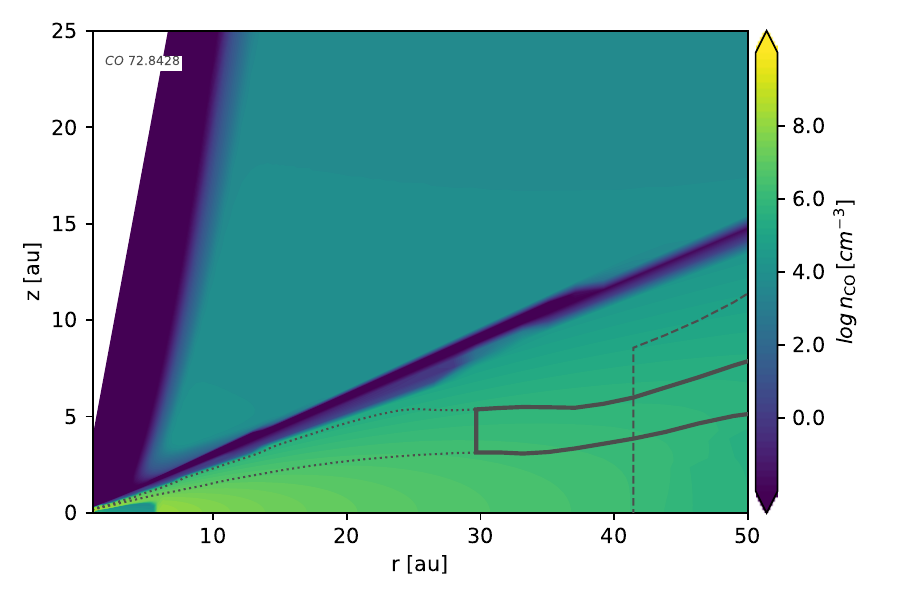}
    \includegraphics[width=0.5\textwidth, trim={0cm 0cm 0cm 0}]{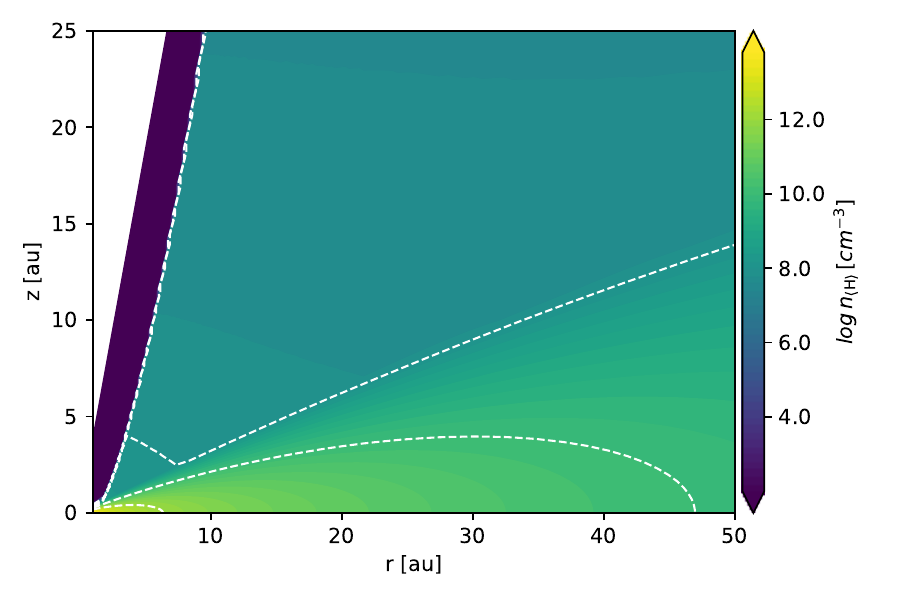} \\
    \includegraphics[width=0.5\textwidth, trim={0cm 0cm 0cm 0}]{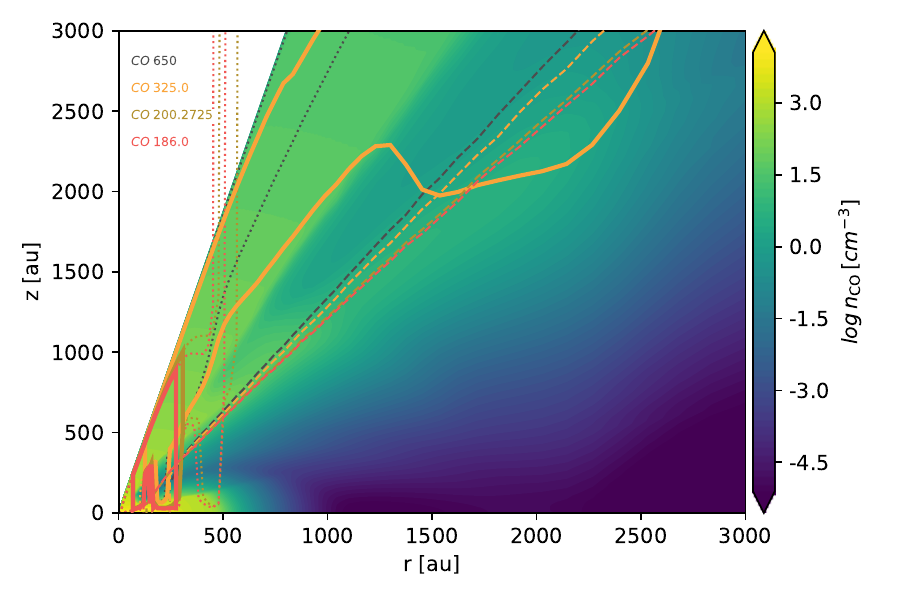}
    \includegraphics[width=0.5\textwidth, trim={0cm 0cm 0cm 0}]{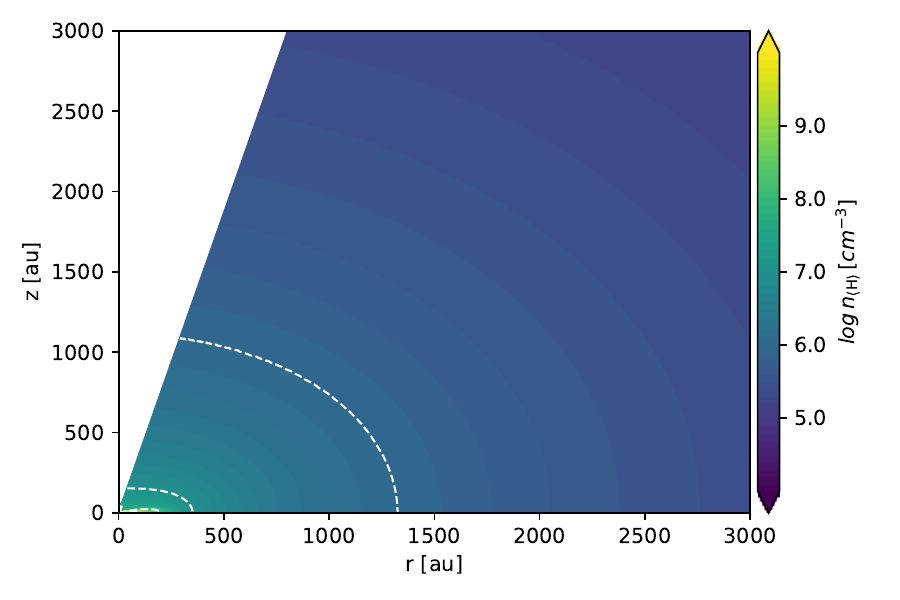} \\
    \includegraphics[width=0.5\textwidth, trim={0cm 0cm 0cm 0}]{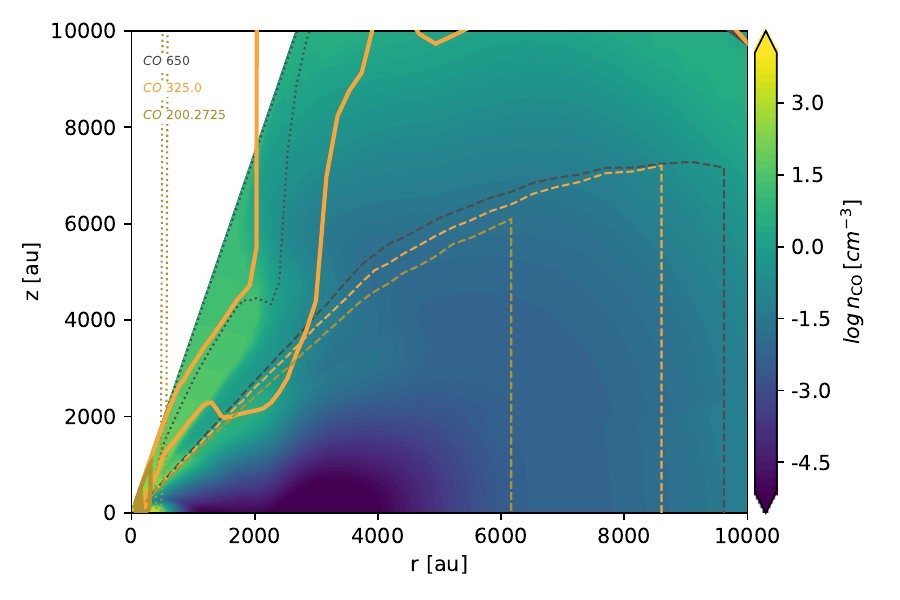}
    \includegraphics[width=0.5\textwidth, trim={0cm 0cm 0cm 0}]{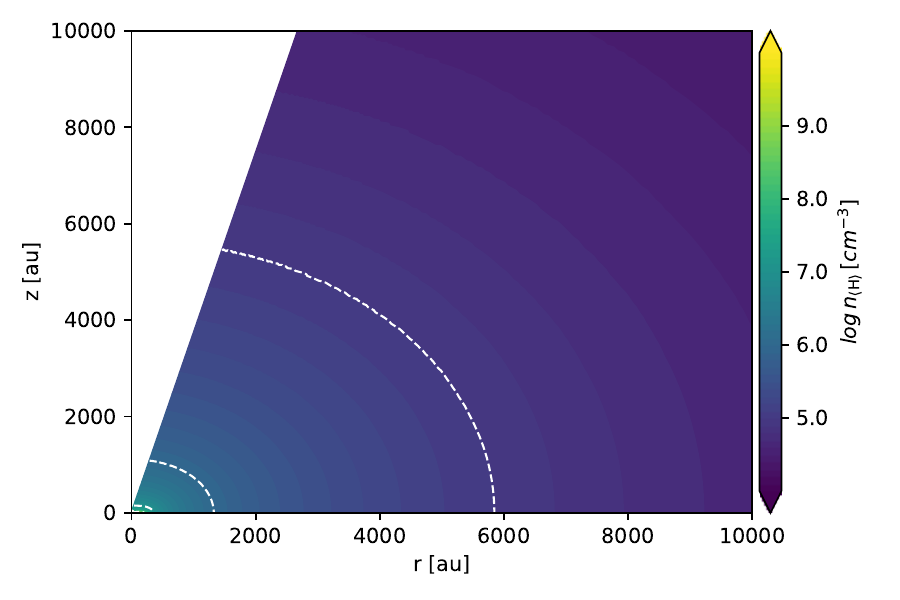}
        \caption{
            \footnotesize
             Spatial distribution of the number density of CO and Hydrogen. The first line shows the smallest scale where the accretion disk is dominating.
             For the CO emission, there is a gap between the disk and the envelope which extends to a radius of about 37 au at a height of 11 au. This gap is
             caused by numerical effects at the transition region. The second and third line show plots up to 3000 and 10000 au distance from the central object,
             where the disk does not play a significant role any more. The dashed lines show the level of the continuum emission at the respective wavelength. The
             dotted line shows the spatial region where 15-85 \% of the emission in vertical direction originates from. The box shows the spatial region where
             15-85 \% of the emission in radial direction are originating from.
        }
        \label{CO_H_density}
\end{figure*}

\begin{figure*}[htb]
    \includegraphics[width=0.5\textwidth, trim={0cm 0cm 0cm 0}]{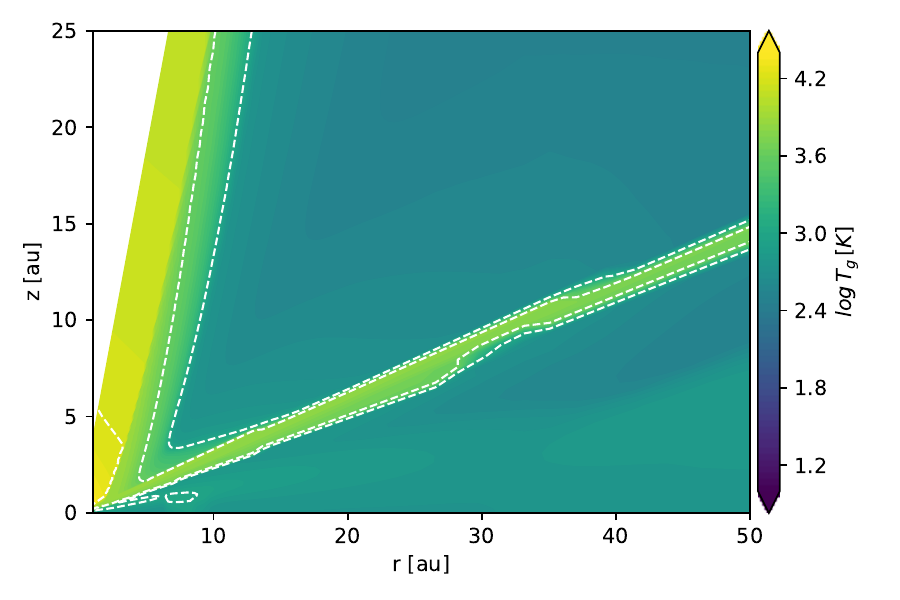}
    \includegraphics[width=0.5\textwidth, trim={0cm 0cm 0cm 0}]{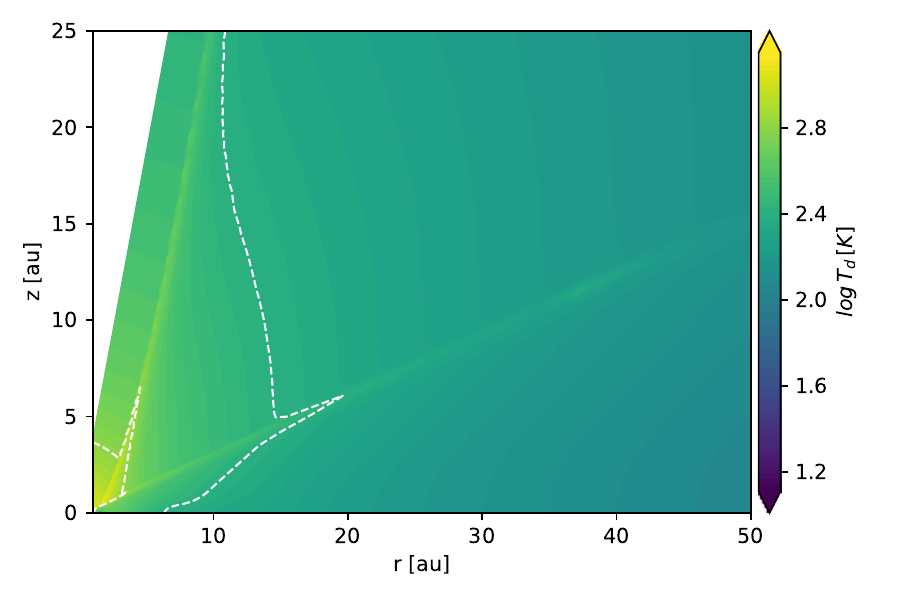}
    \includegraphics[width=0.5\textwidth, trim={0cm 0cm 0cm 0}]{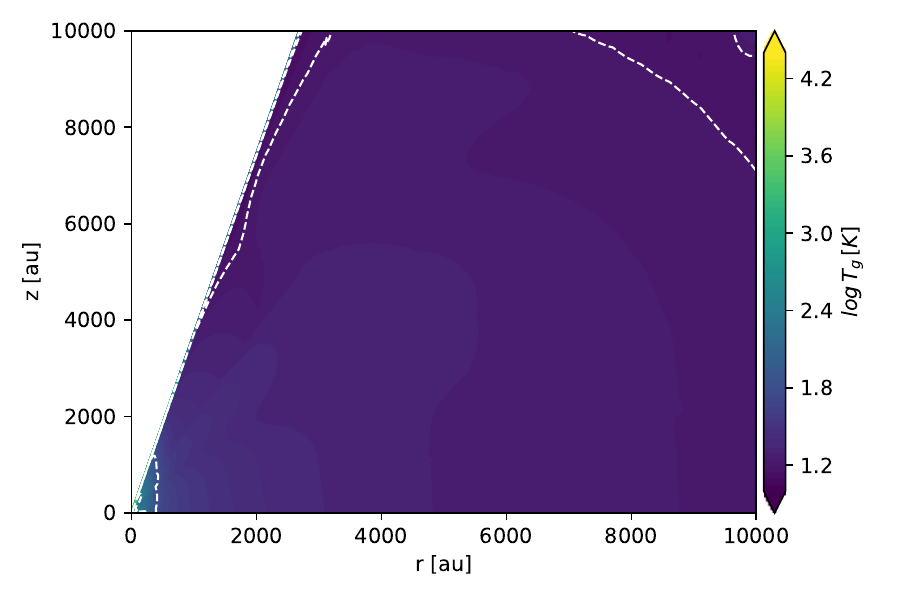}
    \includegraphics[width=0.5\textwidth, trim={0cm 0cm 0cm 0}]{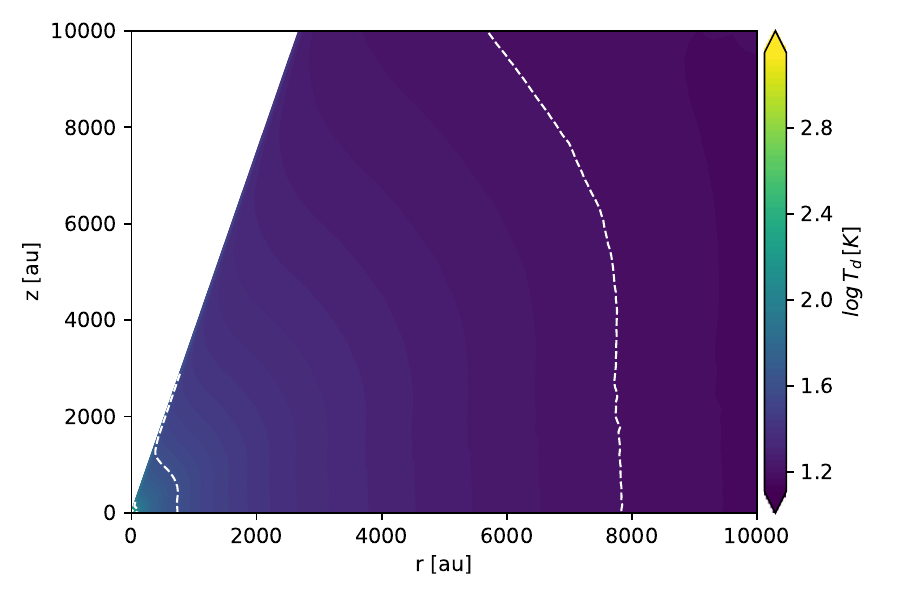}
        \caption{
            \footnotesize
            Spatial variation of the gas component (left) and dust (right) temperature. The highest temperatures of the gas are reached on the outflow cavity on
            small distance to the central source. The
            hot area between the disk and envelope is caused by numerical effects due to the reduced number density of CO in this region (see Fig. \ref{CO_H_density}).
        }
        \label{tgas}
        \label{tdust}
\end{figure*}

        Fig. \ref{CO_H_density} shows that the number density of CO is particularly high in the envelope close to the outflow cavity
        (panels with spatial extent up to 3000 au) and in the disk (panelss with spatial extent up to 50 au). 
        A very high temperature of the gas causes a dissociation and
        depopulation of the CO (see Fig. \ref{tgas}) which consequently leads to a reduced emission from that region. Too low temperature, however, does not
        result in the strong emission lines we see from the Herschel observation, as the short lines would be then too faint.
        ProDiMo allows us to pinpoint down the origin of the
        different CO lines, which is also shown in Fig. \ref{CO_H_density}. While the short-wavelength lines like the 72.8 \textmu m emission still originates
        predominantly from the disk,
        the emission at 186 \textmu m already comes mainly from the envelope close to the cavity, and emissions at $\geq$325 \textmu m have their
        origin only in the envelope.
        We therefore deduce that Re 50 N IRS 1 must consist of a relatively complex set of temperatures, caused by multiple heating processes which
        are spread over a larger spatial area.
        In the Fig. \ref{CO_H_density}, the level of the continuum emission at the respective wavelength (visible in the top left corner of the panels with CO emission, in micrometer) is shown, in addition to the area where most of the emission in vertical and radial direction is created. An important result for observations at these specific wavelengths. The CO density is strongly reduced in the envelope in the line of sight with the disk and the central source, up to a distance of $\approx 4300~au$. This effect is caused by the freeze-out of the CO at low temperature with a weaker effect at low densities (i.e. larger distance). Closer to the outflow cavity, photo-desorption occurs as the UV radiation can penetrate into then envelope, which is however blocked by the disk. The behaviour of the CO abundance structure is as expected, see \cite{Rab2017} for more details.

    \subsection{Rotational diagrams} \label{subsection:rotational_diagrams}
\begin{figure}
    \includegraphics[width=0.5\textwidth, trim={0cm 0cm 0cm 0}]{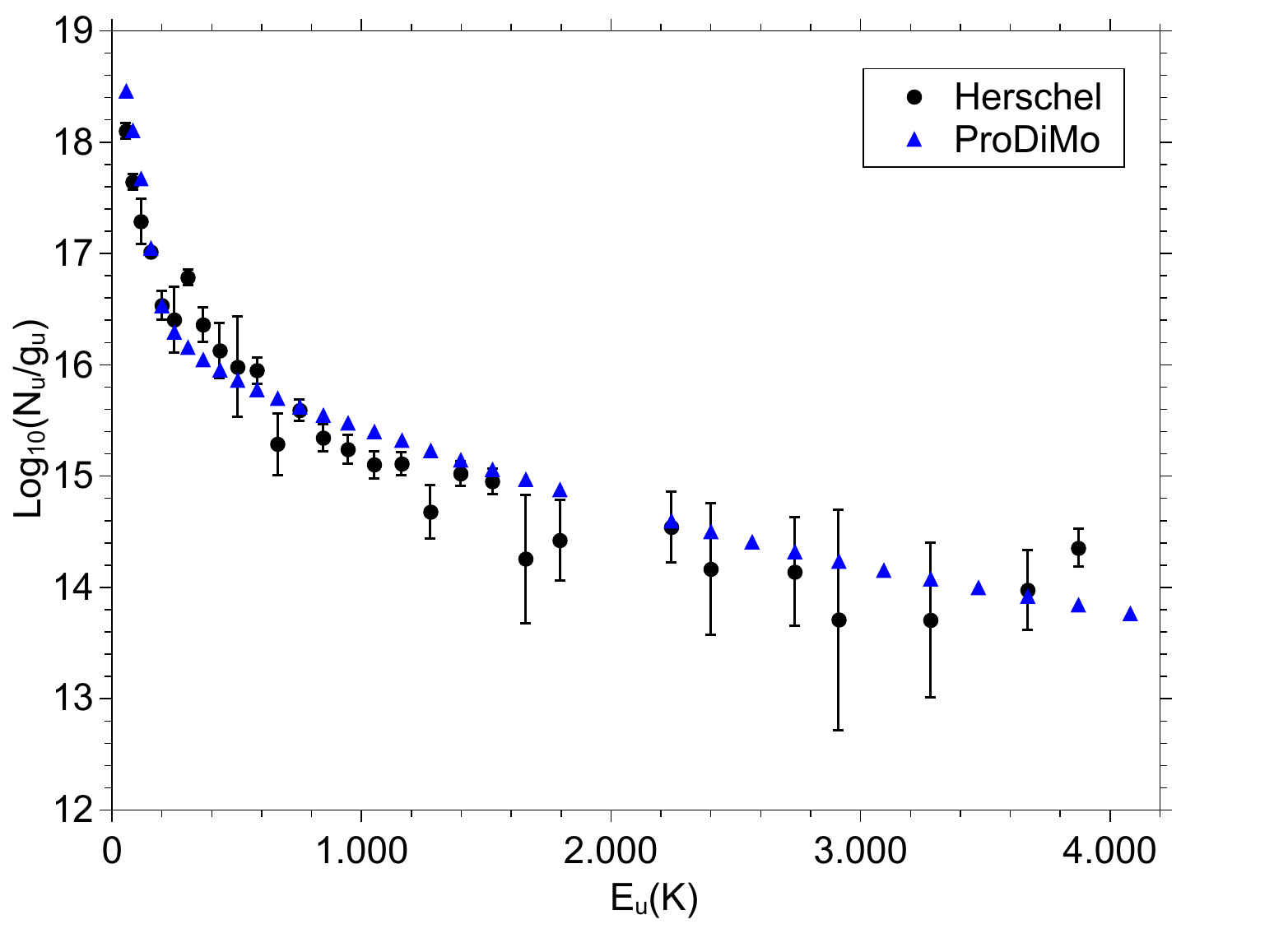}
        \caption{
            \footnotesize
            CO rotational diagram for the best model compared to the observational data obtained with Herschel. The gap around 2000 K is caused by a
            non-detection of the lines in the Herschel data.
        }
        \label{rotatinaldiag}
\end{figure}

        Fig. \ref{rotatinaldiag} shows the rotational diagram for CO emission for our best model and the Herschel observation of the object. This figure is an
        alternative way to show the line emission from Fig. \ref{emissionlines}, but in a normalized way. The diagram can be used to obtain the excitation
        temperatures by measuring the slope
        of the curve, and column density of the gas by measuring the offset for observational dat
        (see \citealt{GoldsmithLanger}, \citealt{GreenRotDiag}, \citealt{Dionatos2013}). The figure shows that the slope, i.e. the temperature distribution of the curve
        is matched by the model. However, we can see that the column density of the model is slightly higher than what we observed. The approach to split the emission
        lines into three domains with assumed LTE from \cite{Postel_2019} turns out to be a solid approximation, albeit used due to the different Herschel
        instruments (PACS, SPIRE SSW, SPIRE SLW).

    \subsection{UV field and accretion}
        As we modelled the condition of Re 50 N IRS 1 during the phase of higher activity, it is possible that the UV field and the accretion rate are lower now.
        As ProDiMo does not have the capabilities to model directly the effects of shocks leading to sufficiently high temperatures in certain areas, especially along the outflow cavity,
        we bypass this with the introduction of the UV field originating from the central source.
        For reaching sufficiently high temperatures for the CO transitions with short wavelengths and the [\ion{O}{i}], the respective parameters appear to be inevitable
        in the constraints of ProDiMo. It turns
        out that the UV field and the disk accretion are both a necessity to create the short emission lines in the observed combination, and the accretion of the
        envelope is the main parameter in our model, responsible for the overall shape of the SED for this object.

\section{Discussion}
    The major difficulty is the distribution of the temperature around the source, which cannot only be provided by the central source.    
    The matching of the emission lines shows that a simple envelope + disk model of Re 50 N IRS 1 is not sufficient, but the system requires a higher
    complexity of additional heating mechanisms. We therefore use the UV field to model the effects of shocks in jets/outflows to recreate the high-J CO lines,
    since ProDiMo does not have the capabilities to
    model them directly. Evidence for the presence of a protostellar jet is provided by the strong [\ion{O}{i}] lines in the Herschel spectra and was shown in
    other works to be a source for a radiation component (see \citealt{2017A&A...597A..64D} and \citealt{2020arXiv200601087D}).
    The inclination
    and the opening angle of the outflow cavity
    of the object play a main role in the visibility of the gas emission lines, as they affect the visibility of the accretion disk, where a large fraction
    of the emission originates from. The envelope of Re 50 N IRS 1 is likely still dense enough to absorb most of the emission lines in the line of sight
    of the disk. Most of the CO emission originates in the inner disk, at some tens of au, where high J transition lines in our model originate from. For the
    intermediate J
    transition lines, the envelope close to the outflow cavity turns out to be the major source, where another high population of CO is located. The cavity is
    also where the highest temperatures of the gas occur, which is also where the radiation field of shocks would be created at. Both, the CO abundance and temperature distribution is in agreement with former works on outbursts,
    like the hydrodynamical simulations from \cite{2013A&A...557A..35V} which targeted the behaviour of CO.
    
    For some of the spectral features
    in the SED, e.g. the narrow absorption of CO$_2$ ice at 15 \textmu m, no matching happened since ProDiMo currently does not have the capabilities to model
    these features.
    
    The high accretion rate of the circumstellar disk of 6 $\times 10^{-4}$ $\mathrm{M}_\sun$ yr$^{-1}$ would cause the accretion disk to dissipate, if it would stay at this high level for a long time, since the mass infall rate of the envelope with 1.35 $\times 10^{-5}$ $\mathrm{M}_\sun$ yr$^{-1}$ which feeds the disk over a long time period
    is significantly lower. This means that the object cannot keep the current properties but the disk accretion is currently in a burst phase which wil
    last for a limited amount of time before a lowering of the disk accretion rate must start.
    The Spitzer data have been observed some time before the Herschel data (2010 vs. 2013), so during the brightening of the object the SED in the near- and mid-IR could
    have changed. A check with recent photometry or even a spectrum would be interesting to confirm this. For a possible observation, a major part of the SED in the near-
    and mid-IR should be traced while the spectral resolution plays a minor role.
    
    When calculating the stellar radius based on the values of the modelled object via $R_{star} = 0.5 \dot{M} G M_{star} / L_{accr}$ (assuming that half of the
    energy goes into the star and half escapes), we would end up with a stellar radius of 52 $R_\sun$ during the outburst, a too large value compared to models. This equation also assumes that the luminosity is dominated by the accretion luminosity and is therefore a lower limit. Given the small impact on
    the SED, we think that a reduction of the stellar mass to 0.3 $\mathrm{M}_\sun$ is feasible, leading to a reduced radius of 31 $R_\sun$ which appears still high. A larger distance than the used 414~pc would lead to higher luminosities, and thus, a lower radius. A 10\% increase ($\approx$ 450 pc) would lead to 110 $\mathrm{L}_\sun$, and a radius of less than 26 $R_\sun$, which is more of a reasonable value according to recent numerical simulations of stellar characteristics during accretion bursts \citep{2019MNRAS.484..146E}. \citet{Grossschedl_2018} showed that the Orion A cloud is inclined from the plane of sky and reaches up to 470 pc, depending on the Galactic longitude. Based on their Figure 3, at $l=211.6^\circ$, YSOs generally are have distances of 425-450 pc (the range is consistent with their derived median distance for the Tail of Orion A, L1641, $\approx 430$ pc, although they derive an average distance of $409.36 \pm 31.9$ for the above longitude, using a 1-degree longitude bin, albeit with a large uncertainty), suggesting that a higher distance for Re 50 N IRS 1 is possible.
    
    For some parameters, there seem to be strong differences among other works.
    \cite{Gramajo_2014} listed for the object a distance of 460 pc and a luminosity of $50~\mathrm{L}_\sun$ while \cite{2015ApJ...805...54C} came to a luminosity of $250~\mathrm{L}_\sun$. Also other parameters in their model, like the accretion rate, differ significantly from our results. As we mentioned earlier, \cite{2015ApJ...805...54C} assume that a dust clearing effect occured, based on near-IR data. However, we see from mid-IR to sub-mm data that there must still be a very strong accretion, possibly in combination with the temporal dust clearing effect at shorter wavelengths. This could also be a reason for the higher optical extinction and variation they mention, ranging from $A_V=50-26~mag$.
    \cite{2018MNRAS.474.4347C} derives an inclination of $70^\circ$, a result that is in disagreement with our model. Based on the Spitzer data and as we show in Fig. \ref{SED_grid10}, the inclination in our model is not expected to be above $22.5^\circ$ while $15.3^\circ$ reproduce the observational data in the best way.
    
    Assuming that the envelope temperature is in the order of 10-20 K, most models of FUors infer a lower envelope infall rate (a few $10^{-6} \mathrm{M}_\sun/yr$) if the
    core was initially close to a Bonnor-Ebert sphere. The significantly higher infall rate therefore means that the temperature of the core is much higher, possibly
    due to a UV source, or the initial pre-stellar core was off the equilibrium. The latter could happen if the core was created by a shock wave.

    \citet{Postel_2019} include several objects with a similar shape of the SED as Re 50 N IRS 1, e.g. PP 13 S,
    V346 Nor, V883 Ori and HH 381 IRS. However, the emission lines among these objects appear very different, raising the question of the reason for this.
    From our results with ProDiMo for Re 50 N IRS 1, it
    appears that these objects, if they show similar line properties, namely PP 13 S and V346 Nor, must host a similar distribution of the temperature in the disk and
    outflow cavity. Thus, objects like V883 Ori and HH 381 IRS, which show very faint emission lines, could consist of a colder disk, which would imply that
    the activity of these objects is way lower than the one of Re 50 N IRS 1. Also, inclination could play a role for these objects, with an envelope shadowing emission lines from the disk, as proposed by \cite{Green2006}.
    There could be also a reduced outflow cavity, obscuring the disk. This would mean that they are in an earlier evolutionary stage if the cavity did not
    become large yet, but the envelope shows a similar mass infall rate as later objects.
    Objects with a similar SED as Re 50 N IRS 1 would be of interest for further modelling with ProDiMo, to test these assumptions, and also multi-dish (sub-)mm
    observations with high angular resolution would provide valuable information about the inner regions of the bjects, in particular the outflow cavity and the disk,
    which have not yet been observed, especially if they can provide information about the temperature.
    
    While the observed features of Re 50 N IRS 1 could be well reproduced in this work, some line features in other objects could be an interesting
    challenge. HH 354 IRS shows a different shape of the SED, which indicates another evolutionary stage. Informative here would be to check the connection
    of the SED with the presence of the strong oxygen emission at 63 \textmu m and the relatively weak emission of CO and oxygen at 145.5 \textmu m.

\section{Summary}
    We provide a ProDiMo model of the young stellar object Re 50 N IRS 1 / HBC 494 which matches observational data of Herschel and Spitzer during a phase of
    re-brightening of the object, covering both the SED and emission lines of CO, $^{13}$CO and two [O I] lines.
    We find a way to recreate the gas lines of the target by an accreting envelope and a
    circumstellar disk with high accretion rate and an additional UV field to heat up the disk and outflow cavity to form the necessary environmental conditions.
    The gas lines appear to have their origin in the disk and the transition region of the envelope and the outflow cavity, created by shocks in jets/outflows.
    The envelope turns out to be dense enough to absorb most emission in the line of sight to the disk. The system requires episodic accretion to be consistent
    over long time with a high disk accretion rate and a significantly smaller mass infall rate of the envelope.

\begin{acknowledgements}
    We thank A. Kospal for providing the ALMA data for Re 50.
\end{acknowledgements}

\bibliographystyle{aa}
\bibliography{./publication_1}

\begin{appendix}

\section{Disk accretion heating} \label{appendix:diskaccretionheating}
For the viscous heating we use the approach of \citep{1998ApJ...500..411D} and adapted it to account for the tapered outer edge of the disk.
\begin{equation}
    F_{\rm vis} (r) = \frac{3 G M_{\rm *} \dot{M}}{8 \pi r^3}\cdot \left( 1-\sqrt{\frac{R_*}{r}}\right)\cdot
    \exp{\left(-\left(\frac{r}{R_\mathrm{tap}}\right)^{(2-\epsilon)}\right)}   \,\,\,\,\, [\textrm{erg cm}^{-2}  \textrm{s}^{-1}],
\end{equation}
where $r$ is the distance to the star, $G$ is the gravitational constant, $M_{\rm *}$ and $M_{\rm *}$ the stellar mass and radius, $R_\mathrm{tap}$ the disk tapering-off radius and $\epsilon$ the disk column density power index. To distribute the heating per unit surface over the disk heat we use 
\begin{equation}
    \Gamma_{\rm vis}(r,z) =  F_{\rm vis} (r) \frac{\rho^P(r,z)}{\int \rho^P(r,z') dz'} \,\,\,\,\,[\textrm{erg cm}^{-3} \textrm{s}^{-1}] ,
\end{equation}
where $\rho$ is the gas density and $P=1.25$ to avoid unstoppable heating in the upper disk layers where $\rho$ drops towards zero.

\end{appendix}

\end{document}